\documentclass[aps,prl,twocolumn,showpacs,amsmath,amssym,floatfix, superscriptaddress,nofootinbib]{revtex4-1}
\usepackage{graphicx}

\newcommand {\be}{\begin{equation}}
\newcommand {\ee}{\end{equation}}

\usepackage{xcolor}

\newcommand{\red}{\textcolor{red}}

\begin{document}

\title{Intermittent chaotic chimeras for coupled rotators}
\date{\today}

\author{Simona Olmi}
\email{simona.olmi@fi.isc.cnr.it}
\affiliation{CNR - Consiglio Nazionale delle Ricerche - Istituto dei Sistemi 
Complessi, via Madonna del Piano 10, I-50019 Sesto Fiorentino, Italy}
\affiliation{INFN Sez. Firenze, via Sansone, 1 - I-50019 Sesto Fiorentino, Italy}

\author{Erik A. Martens}
\email{erik.martens@sund.ku.dk}
\affiliation{Department of Biomedical Sciences, University of Copenhagen, 
Blegdamsvej 3, 2200 Copenhagen, Denmark}
\affiliation{Department of Mathematical Sciences, University of Copenhagen, Universitetsparken 5,
2100 Copenhagen, Denmark}
\affiliation{Max Planck Institute for Dynamics and Self-Organization, 37077 G\"ottingen, Germany}

\author{Shashi Thutupalli}
\email{shashi@princeton.edu}
\affiliation{Dept. of Physics, Princeton University, Princeton, New Jersey 08544,  USA}
\affiliation{Dept. of Mechanical and Aerospace Engineering, Princeton University, Princeton, New Jersey 08544,  USA}

\author{Alessandro Torcini}
\email{alessandro.torcini@cnr.it}
\affiliation{CNR - Consiglio Nazionale delle Ricerche - Istituto dei Sistemi 
Complessi, via Madonna del Piano 10, I-50019 Sesto Fiorentino, Italy}
\affiliation{INFN Sez. Firenze, via Sansone, 1 - I-50019 Sesto Fiorentino, Italy}

\begin{abstract}
Two symmetrically coupled populations of $N$ oscillators with inertia $m$ display chaotic solutions with broken symmetry similar to experimental observations with mechanical pendula. In particular, we report the first evidence of {\it intermittent chaotic chimeras}, where one population is synchronized and the other jumps erratically between laminar and turbulent phases. These states have finite life-times diverging as a power-law with $N$ and $m$. Lyapunov analyses reveal chaotic properties in quantitative
agreement with theoretical predictions for globally coupled dissipative systems.
\end{abstract}


\maketitle
Chimera states are remarkable dynamical states emerging in populations of coupled identical oscillators, where the population splits into two parts: one synchronized and the other composed of incoherently oscillating elements~\cite{Kuramoto2002}. These states have been initially discovered in chains of nonlocally coupled oscillators, however they can equally emerge in models of globally coupled populations~\cite{Kuramoto2002, abrams2004,abrams2008}. Chimeras have been observed in a  repertoire of different models~\cite{abrams2004,Shima2004,abrams2008,Martens2010bistable,Omelchenko2008, Pikovsky2008,Laing2009,Olmi2010}, and in various experimental settings, including mechanical~\cite{MartensThutupalli2013, kapitaniak2014}, 
(electro-)chemical ~\cite{Wickramasinghe2013,Tinsley2012} and lasing systems~\cite{Hagerstrom2012}, among others. Usually, the incoherent oscillators give rise to regular macroscopic dynamics which are either stationary, periodic (so-called breathing chimera) or even quasi-periodic~\cite{abrams2008,Pikovsky2008}.  
Only recently, spatio-temporally chaotic chimeras have been numerically identified
in rings of coupled oscillators~\cite{bordyugov2010,wolfrum2011,omelchenko2011loss, sethia2014}.
However, a detailed characterization of the dynamical properties of these states have been reported only for phase oscillators with finite-range interactions: in this case chimeras are transient, and {\it weakly chaotic}~\cite{wolfrum2011,wolfrum2011spectral}. More specifically, the life-times of these states diverge exponentially with the system size, while their dynamics becomes regular in the thermodynamic limit. In contrast, for globally connected populations, chaotic chimeras have so far only been observed in pulse-coupled oscillators~\cite{pazo2014} without further analysis.
 
\begin{figure}[h]
\begin{center}
\includegraphics*[angle=0,height=5cm,width=8.5cm]{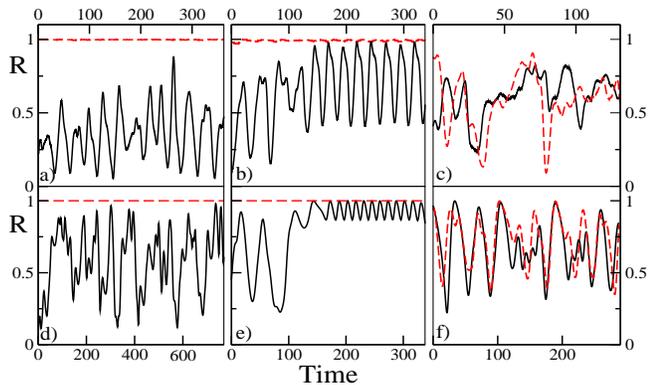}
\end{center}
\caption{(color online) Order parameters $R^{(1)}$ (solid black curve)
and $R^{(2)}$ (dashed red curve) for the two coupled populations versus time.
(a-c) Experimental measurements; (d-f) numerical simulations of the model (\ref{eq1})
with $m=10$. Initial conditions are BSCs for (a,b,d,e) and UCs for (c,f) and $N=15$. \red{The experiments in panels (a-b) (panel (c)) are carried out with $f = 160$~beats per minute (bpm) and  $l= 17$cm ($f = 184$~bpm and $l= 25$ cm) and experimental time is measured in seconds.}
}
\label{fig.1}
\end{figure}

In this Rapid Communication, we report the existence of various irregular solutions with broken 
symmetry in an experiment with two mechanically coupled populations of pendula 
and analyze in depth a model which reproduces this complex dynamics. Three pertinent examples from the experiment, shown in Fig.~\ref{fig.1} (a-c), are of particular interest. The first two examples represent chimeras: in Fig.~\ref{fig.1} (a), the order parameter of the desynchronized population oscillates quite irregularly, while in Fig.~\ref{fig.1} (b) it enters a regime of almost periodic oscillations. Fig.~\ref{fig.1} (c) reports a situation where both populations are irregularly oscillating. We introduce a simple model (Eq.~\eqref{eq1}) capable of reproducing all these different dynamical behaviors, as it can be appreciated by the simulations reported in Figs.~\ref{fig.1} (d-f). The model consists of two symmetrically globally coupled populations of $N$ Kuramoto phase oscillators with inertia. 

The introduction of inertia allows the oscillators to synchronize via the adaptation 
of their own frequencies, in analogy with the mechanism observed in certain species of fireflies~\cite{ermentrout1991}. The modification of the classical Kuramoto model with the 
addition of an inertial term leads to first order synchronization transitions and complex hysteretic phenomena~\cite{tanaka1997first,tanaka1997self, Ji2013, gupta2014, komarov2014, olmi2014}. Furthermore, networks of phase coupled oscillators with inertia have recently been employed to investigate self-synchronization in power grids~\cite{salam1984,filatrella2008,rohden2012, motter2013} and in disordered arrays of Josephson junctions~\cite{trees2005}.
In absence of dissipation, Eq.~\eqref{eq1} reduces to the Hamiltonian mean-field model -- a paradigm of long-range interacting systems~\cite{campa2014_book}. 
 
Our analysis will mainly focus on the solution shown in Fig.~\ref{fig.1} (a), which we 
term as an {\it Intermittent Chaotic Chimera} (ICC). This state exhibits turbulent phases 
interrupted by laminar regimes, analogous to the one reported in 
Fig.~\ref{fig.1} (b). The third state shown in Fig.~\ref{fig.1} (c)
is a {\it Chaotic Two Populations} (C2P) state, here the erratic dynamics is induced
by the evolution of the non-clustered oscillators belonging to both populations.
In particular, we show that ICCs are transient states for finite inertia and system size,
whose life-times diverge as a power-law with $N$ and $m$.
Furthermore, in the thermodynamic limit the intermittent oscillations disappear and
the turbulent regime prevails over the laminar one. 
The stability properties of the ICC can be ascribed to the {\it universality class} of globally coupled systems~\cite{takeuchi2011}, which are distinct from those reported for chaotic chimeras in chains of oscillators~\cite{wolfrum2011spectral}. This result clearly illustrates that the stability of chimera states strongly depends on the underlying network topology.

\begin{figure}[ht]
\includegraphics*[angle=0,width=0.4\textwidth,height=0.4\textwidth]{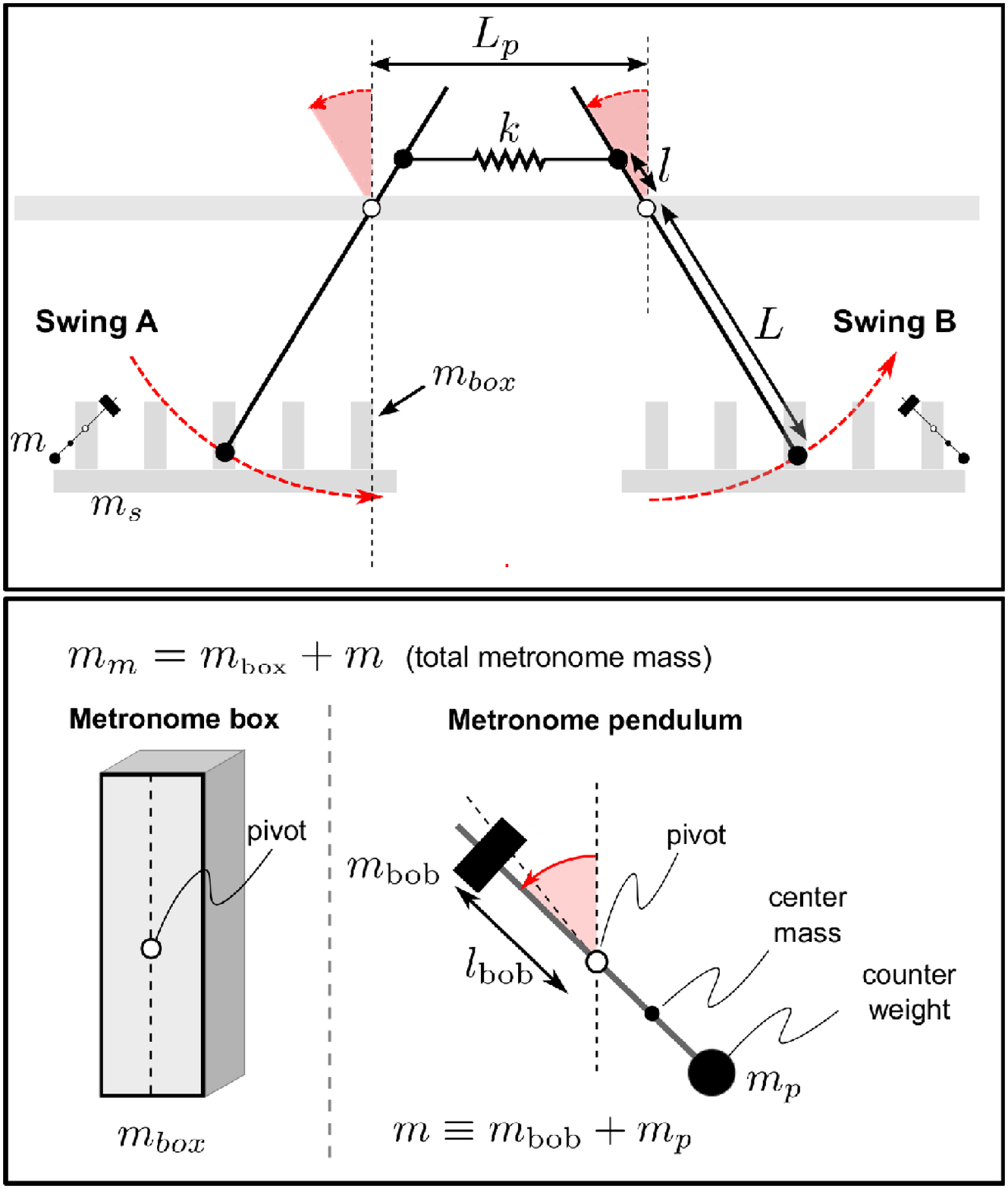}
\caption{\label{fig:swingsystem} \red{
(color online) Experimental setup: A sketch of the the mechanical system studied in~\cite{MartensThutupalli2013}, 
it is composed of two swings (A,B) coupled with a spring mechanism, each of which is loaded  with $N=15$ metronomes.}
}
\end{figure}

\red{
\textit{Experimental setup.} 
The setup is shown in Fig.~\ref{fig:swingsystem} and it is 
identical to the one described in~\cite{MartensThutupalli2013}.
The experiment is composed of purely mechanical parts -- in particular, following
\cite{abrams2008}, two populations of non-locally coupled metronomes were considered: within each population oscillators were coupled strongly, but were coupled more weakly to the neighboring population. Metronomes take the role of self-sustained oscillators~\cite{Pantaleone2002}, whose working principle is identical to Huygens' pendulum clocks~\cite{Bennett2002}, except that the escapement in the metronome is driven by a spring rather than a mass pulled by gravity. While friction inherent to the mechanical elements attenuates large-amplitude oscillations toward the unperturbed amplitude associated with its unperturbed frequency, small pendulum oscillations are amplified by the spring that drives the metronome via its escapement mechanism. This lends the metronome the characteristics of a self-sustained oscillator~\cite{Pantaleone2002}. $N=15$ identical metronomes running with identical frequencies were placed on each of two aluminum swings suspended by four rods. The strong coupling within one population is mediated by the motion of the swing onto which the metronomes are attached. As one increases the common frequency $f$ of the metronomes, more momentum is transferred to the swing, leading to a stronger coupling among the metronomes. A single swing follows a phase transition from a disordered to a synchronized state as the coupling within the 
population increases~\cite{Pantaleone2002}. 
}
\red{
The weaker coupling between the two swings is facilitated by a pair of tunable steel springs, attached to the adjacent rods of the opposing swings. The distance of the spring relative to the pivot can be adjusted: this changes the spring lever $l$ and the associated torque, thus effectively tuning the spring coupling strength between the two metronome populations. 
}

\red{
The motion of swings and metronomes is visualized by attaching UV fluorescent spots on swings and metronome pendula, and the spot motion is digitally recorded using a DSLR photocamera. Subsequently digital data analysis is 
used to measure the swing and pendulum motion. 
}
\red{
The relative motion of the pendula is reconstructed by subtraction of the swing coordinates. Amplitudes and phases are then obtained via Hilbert transformation of the signal. The phases are used to quantify the level of synchronization for each population
by using the order parameters defined in the following. For exhaustive details on the experimental setup  and methods, see~\cite{MartensThutupalli2013}.}

\red{\textit{Model and Methods}.} We consider a network of two symmetrically coupled populations of $N$ oscillators. The phase $\theta_i^{(\sigma)}$ of the $i$-th oscillator in population $\sigma=1,2$ evolves according to the 
differential equation
\begin{equation}
\label{eq1} 
m\ddot{\theta}_i^{(\sigma)} + \dot{\theta}_i^{(\sigma)}=\Omega+\sum_{\sigma'=1}^2 
\frac{K_{\sigma\sigma'}}{N} \sum_{j=1}^N  \sin{\left(\theta_j^{(\sigma')}-\theta_i^{(\sigma)}-\gamma\right)}
\end{equation}
where the oscillators are assumed to be identical with inertia $m$, 
natural frequency $\Omega=1$ and a fixed phase lag $\gamma= \pi - 0.02$. 
The self- (cross-) coupling among oscillators belonging to the same population 
(to different populations) is defined as $K_{\sigma\sigma} = 0.3$ ($K_{\sigma\sigma'} \equiv K_{\sigma'\sigma} = 0.2$), with $K_{\sigma\sigma} > K_{\sigma\sigma'}$ as in  previous studies on chimera states~\cite{abrams2008,montbrio2004}.
We consider two types of initial conditions:
i) broken symmetry conditions (BSCs), realized by initializing the first (second) population 
with identical (random) phases and frequencies, which may lead to the emergence of chimera states, and
ii) uniform conditions (UCs) where both populations are initialized with random values, and can result in a C2P state~\cite{note2}.

The collective evolution of each population will be characterized in terms of the
macroscopic fields $\rho^{(\sigma)}(t)=R^{(\sigma)}(t) \exp{[i \Psi(t)]} = N^{-1}\sum_{j=1}^N 
\exp{[i\theta^{(\sigma)}_j(t)]}$. The modulus $R^{(\sigma)}$ is an order parameter 
for the synchronization transition being one (${\cal O}(N^{-1/2})$) for synchronous 
(asynchronous) states. 

\red{The microscopic stability can be measured
in terms of the associated ordered spectrum of the Lyapunov exponents (LEs) $\{ \lambda_i \} \enskip i=1,\dots, 4 N$, representing the exponential growth rates of infinitesimal perturbations. 
The dynamics is chaotic whenever the maximal LE $\lambda_M \equiv \lambda_1$ is positive.
For the studied model, presenting a constant viscous dissipative term, 
the spectrum satisfies the following pairing rule~\cite{dressler1988} :
$\lambda_i+\lambda_{4N-i+1}=-1/m$.
Therefore the analysis can be limited to the first $2 N$ exponents.
The Lyapunov spectrum can be numerically estimated 
by employing the standard method reported in~\cite{benettin1980}. This amounts to
consider for each LE $\lambda_k$, the evolution of a 4$N$-dimensional tangent vector
$\mathcal{T}^{(k)} = \{\delta\dot{\theta}_i^{(1)},\delta\dot{\theta}_{i}^{(2)},\delta{\theta}_i^{(1)},\delta{\theta}_{i}^{(2)}\} \quad i=1,\dots,N$, whose dynamics is ruled by the linearization of Eq.~(\ref{eq1}):
\begin{equation}
\label{eq2}
m \delta\ddot{\theta}_i^{(\sigma)} + \delta\dot{\theta}_i^{(\sigma)}= \sum_{\sigma'=1}^2 \frac{K_{\sigma\sigma'}}{N} \sum_{j=1}^N 
A_{ji}^{ ( \sigma' \sigma ) }
(\delta\theta_j^{(\sigma')}-\delta\theta_i^{(\sigma)})
\end{equation}
where 
$A_{ji}^{(\sigma' \sigma)} = \cos{ (\theta_j^{(\sigma')}-\theta_i^{(\sigma)}-\gamma)}$.
The orbit and the tangent vectors is followed for a time lapse $T_s$ by 
performing Gram-Schmidt ortho-normalization at fixed time intervals $\Delta t$, 
after discarding an initial transient evolution $T_t$. We have employed $\Delta t=5$ and  $T_t = 5,000$, for BSCs we have integrated the system for times $8 \times 10^4 \le T_s \le 3 \times 10^5$ with $N = 100, \dots, 800$ and for UCs for times $3 \times 10^4 \le T_s \le 1 \times 10^6$ with $N = 100, \dots, 400$. The integrations have been 
performed with a $4^{\rm th}$ order Runge-Kutta scheme with time step $5 \times 10^{-4}$.
}

\red{
 A characterization of the dynamical evolution on short time scales can be achieved
by considering the probability distribution function $P(\Lambda)$
of the finite time LE $\Lambda$~\cite{chaos}. The finite time LE is 
calculated by estimating the exponential growth rate of the magnitude of the maximal tangent vector $\mathcal{T}^{(1)}$ over finite time windows $\Delta t$, namely
$ \Lambda =\frac{1}{\Delta t} \ln ||\mathcal{T}_i^{(1)}(\Delta t)||$
where $||\mathcal{T}^{(1)}(0)||  \equiv 1$. In order to estimate $P(\Lambda)$, we have collected $100,000$ data points for each system size, obtained from ten different orbits each of duration $T_s = 100,000$ with $\Delta t = 10$. 
}

\begin{figure}[h]
\begin{center}
\includegraphics*[angle=0,height=5cm,width=8cm]{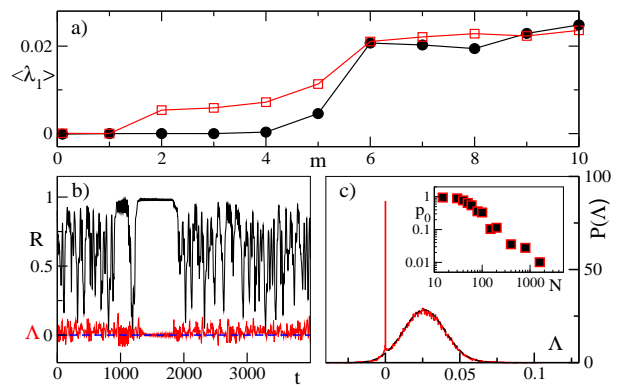}
\end{center}
\caption{(color online) 
(a) Average maximal LE \red{$\langle {\lambda}_1 \rangle$} vs $m$ for $N=100$: 
black circles (red squares) refer to BSCs (UCs).
(b) Order parameter $R$ (black solid line)
and  finite time LE $\Lambda$ (red dashed line) for the chaotic population vs time $t$ for $N=200$ and $m=10$. The corresponding $P(\Lambda)$ (solid red line) is shown in (c) 
with the Gaussian fit (black dashed line). Inset: probability $p_0$ versus $N$.
The  values \red{$\langle {\lambda}_1 \rangle$} are obtained by following each realization for a time span $t=50,000$ and by averaging over 100 different initial conditions.
}
\label{fig.2}
\end{figure}

\textit{Intermittent Chaotic Chimeras}. Starting simulations with BSCs at small masses ($m \le 4$), the system is not chaotic (as shown in Fig.~\ref{fig.2} (a)) and it displays a multitude of coexisting
breathing chimeras~\cite{abrams2008}. Furthermore, while the synchronous state $R^{(1)}= R^{(2)} \equiv 1$ remains stable also in presence of inertia ($m>0$), the stationary chimeras associated to constant order parameters with $R^{(1)} < R^{(2)} \equiv 1$ are not any longer observed. For sufficiently large masses, a solution with broken symmetry emerges, where one population is fully synchronized with $R^{(2)} \equiv 1$, while the other population exhibits wide collective irregular oscillations in the order parameter $R^{(1)} (t)$ between zero and one, as shown in Fig.~\ref{fig.1} (d) and Fig.~\ref{fig.2} (b). These are ICCs and they represent the main subject
of this Rapid Communication.

As shown in Fig.~\ref{fig.2} (b),
the erratic oscillations of $R^{(1)} $ are interrupted by laminar phases, 
where $R^{(1)}$ stays in proximity of one displaying small oscillations.
This regime is characterized by a large part of 
the oscillators in the chaotic population getting entrained to the synchronous population, 
apart a few oscillators, which keep oscillating with distinct identical frequency, 
but with incoherent phases. An analogous regime is also experimentally observed, 
as reported in  Fig.~\ref{fig.1} (b), however due to the smaller size of the populations
the amplitude of the oscillations is larger. Furthermore, by estimating the finite time LE
$\Lambda$, we show that the laminar phases are indeed regular, since they 
are associated to $\Lambda  \simeq 0$ (see Fig.~\ref{fig.2} (b)).
In particular, the distribution $P(\Lambda)$, reported
in Fig.~\ref{fig.2} (c), reveals a clear peak at $\Lambda=0$, associated to the laminar regime, 
superimposed to a seemingly Gaussian distribution.
The probability $p_0$ to observe a laminar phase can be estimated by integrating $P(\Lambda)$ within a narrow interval around $\Lambda = 0$.
This probability is reported in the inset of Fig.~\ref{fig.2} (c) for $10 \le N \le 1600$ and 
it shows a power-law decay with $N$ for sufficiently 
large system sizes, namely $N \ge 50$. This is a clear indication that 
the laminar episodes tend to disappear in the thermodynamic limit.

\begin{figure}[h]
\begin{center}
\includegraphics*[angle=0,height=5cm,width=8cm]{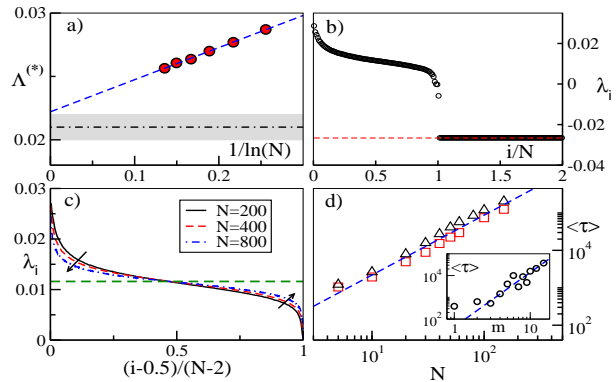}
\end{center}
\caption{(color online) 
(a) $\Lambda^{(*)}$ (symbols) versus  $1/\ln(N)$ for $ 50 \le N \le 1600$ and the corresponding fit, the dot-dashed black curve with the shaded area denotes $\lambda_M(\infty)$ with its error bar. (b) Lyapunov spectrum for $N=100$, \red{the dashed red line is $\lambda^{(s)}$.} (c) Positive part of the Lyapunov spectra for various sizes, the dashed (green) line is $\lambda^{(c)}$. (d) Average life-times \red{$\langle \tau \rangle$} of the ICC vs $N$ for inertia $m=8$ (red squares) and $m=10$ (black triangles). Inset: \red{$\langle \tau \rangle$} vs $m$ for $N=30$ for ICCs. The blue dashed line in the main panel (inset) refers to a power law with exponent 1.60 (2.50). 
The values of \red{$\langle \tau \rangle$} are averaged over 200-3000 different realizations of BSCs and $m=10$ in (a-c).
}
\label{fig.3}
\end{figure}

To characterize the erratic phase, we give an  estimate of the average LE $\Lambda^{(*)}$ restricted to this phase. This estimate has been obtained by evaluating the maximum of $P(\Lambda)$ with a Gaussian fit to the data, once the channels around $\Lambda = 0$ were removed.
The corresponding data, reported in Fig.~\ref{fig.3} (a) for various $N$, reveal a clear decay of $\Lambda^{(*)}$ as $1/\ln{(N)}$. Furthermore, 
the extrapolated value of $\Lambda^{(*)} \simeq 0.022$ 
for $N \to \infty$ is definitely positive, thus indicating that the chaotic state persists in the thermodynamic limit for finite $m$, contrary to what is usually observed for the Kuramoto model 
in~\cite{popovych2005phase,wolfrum2011spectral}. \red{
This logarithmic dependence of the maximal LE $\lambda_M$ with the system size has been 
previously reported for globally coupled networks in~\cite{takeuchi2011}, where, for 
dissipative systems, the authors have shown analytically that
\begin{equation}
\label{eq3} 
\lambda_M(N) = \lambda^{(c)} + \frac{D}{2} + \frac{a}{\ln(N)} + {\cal O}\left(\frac{1}{\ln^2 (N)}\right)
\enskip ;
\end{equation}
where $\lambda^{(c)}$ is the mean field LE obtained by considering an isolated unit of the chaotic population forced by the two fields $\rho^{(1)}$ and $\rho^{(2)}$, $D$ is the diffusion coefficient associated to the fluctuations of $\lambda^{(c)}$~\cite{takeuchi2011}. In particular, $D$ can be measured by 
the scaling of the variance of $\ln d(t)$, where $d(t)$ is the modulus of the Lyapunov vector associated to the single forced unity. Namely, for sufficiently long times
it is expected that 
$\langle [\ln d(t) - \lambda^{(c)} t]^2 \rangle \simeq D t$ with the average $<\cdot>$ performed over many realizations.}
\red{In the present case, we measured $\lambda^{(c)} \simeq 0.0116(5)$ and $D \simeq 0.0180(10)$, thus the expected asymptotic LE should be $\lambda_M(\infty) \simeq 0.021(1)$, which is in good agreement with the previous reported numerical extrapolation as shown in Fig.~\ref{fig.3}(a). This represents the first quantitative verification of the prediction (\ref{eq3}) for a dissipative system with continuous time and in particular for an intermittent dynamics.
}

A more detailed analysis of the stability of this state can be achieved by estimating the Lyapunov spectra for various system sizes: we observe that the spectrum
is composed of a positive part made of $N-2$ exponents and a negative part composed of $N$ exponents, see Fig.~\ref{fig.3} (b). Two exponents are exactly zero: one is always present for systems with continuous time, while the second arises due to the invariance of Eq.~(\ref{eq1}) for uniform phase shifts. The negative part of the spectrum is composed of an isolated LE, quantifying the longitudinal stability of the synchronized population, and $N-1$ identical LEs, which measure the transverse stability of the synchronous solution~\cite{pecora1995}. \red{The value of this negative plateau in the spectrum
coincides with the mean field LE $\lambda^{(s)} = -0.0266(5)$ calculated for an isolated oscillator of the synchronized population, as shown in Fig.~\ref{fig.3} (b).}

Furthermore, the central part of the positive spectrum reveals a tendency to flatten towards the mean field value $\lambda^{(c)}$ associated to the chaotic population for increasing system sizes, see Fig.~\ref{fig.3} (c) for $N=200, 400$ and 800, while the largest and smallest positive Lyapunov exponents tend to split from the rest of the spectrum. This scaling of the Lyapunov spectra has been found to be a general property of fully coupled dynamical systems. In particular, the authors  in~\cite{takeuchi2011,ginelli2011} have shown that in the thermodynamic limit the spectrum becomes asymptotically flat (thus trivially extensive), but this part is sandwiched between subextensive bands containing typically ${\cal O} (\log﻿N)$ exponents scaling as in Eq.~(\ref{eq3}) with $N$. We can safely affirm that the chaotic population in the ICC reveals properties which are typical of fully coupled systems, contrasting 
with the results reported for chaotic chimeras emerging in spatially extended systems~\cite{wolfrum2011,wolfrum2011spectral}.

Let us now examine if the ICCs are transient states; indeed, we observe for different masses that the chaotic chimeras converge to a regular (non-chaotic) state after a transient time $\tau$. This amounts to the fact that the system remains entrapped in a laminar state,
which could be either a fully synchronous regime or a breathing chimera, without returning 
to the turbulent phase. We have measured the average life times  \red{${\langle \tau \rangle}$~\cite{note1}} of the ICCs for two masses, namely $m=8$ and 10, and various system sizes $5 \le N \le 150$. These results are displayed in Fig.~\ref{fig.3} (d). From the figure, it is clear that for $N \ge 10$ one has a power-law divergence of the synchronization time with an exponent $\alpha \simeq 1.60(5)$. The divergence of $\tau$ is directly related to the vanishing of the laminar phases ($p_0 \to 0$) observable for $N \to \infty$. Unfortunately, \red{due to CPU limitations} we cannot explore larger system sizes to verify that this scaling is present over more decades. However, we can safely affirm that these times are not diverging exponentially with $N$ as reported in~\cite{wolfrum2011}. This is a further indication that our phenomenon has a different nature, which is deeply related to the topology presently considered. Indeed, exponentially diverging transients for metastable states have been usually reported in the context of spatially extended systems~\cite{tel2008}, while metastable states with life-time diverging as \red{${\langle \tau \rangle} \propto N^{\alpha}$} -- with $\alpha \simeq 1.7$ -- have been reported for the Hamiltonian version of our model~\cite{yamaguchi2003,gupta2014}. 
Furthermore, we have tested the dependence of \red{${\langle \tau \rangle}$} on the mass, for one system size, namely $N=30$, and we observe that \red{${\langle \tau \rangle}$} is diverging also as a power law of $m$ with an exponent $2.50(5)$ (see inset of Fig.~\ref{fig.3} (d)). It is important to remark that regular chimeras, appearing for $m=0$, are not transient for this topology, as we have numerically verified.
 
\begin{figure}[h]
\begin{center}
\includegraphics*[angle=0,height=5cm,width=8.cm]{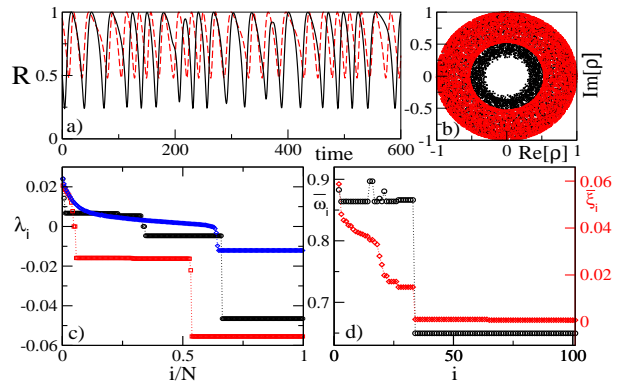}
\end{center}
\caption{(color online) (a) Order parameters $R^{(1)}$ (black solid line) 
and $R^{(2)}$ (red dashed line) vs time; (b) imaginary vs real part of the 
complex fields $\rho^{(1)}$ and $\rho^{(2)}$.
(c) Lyapunov spectra obtained for 3 different realizations of UCs. 
(d) Averaged components $\bar \xi_i$ (red diamonds) together with the corresponding averaged frequencies of the oscillators ${\bar \omega}_i \equiv \frac{\overline{ d \theta_i}}{dt}$ (black circles) for both populations. All the data refer to C2P states and to $m=9$, the system sizes are $N=200$ in (c) and $N=50$ in (a),(b),(d). 
\red{The time averages have been performed a time $T \simeq 5 \times 10^4$.}
}
\label{fig.4}
\end{figure}

\textit{The Chaotic Two Populations State}. With UCs, the system evolves towards chaotic solutions already at smaller masses, namely $m > 1$, as shown in Fig.~\ref{fig.2} (a). With these initial conditions the multistability is strongly enhanced and many different coexisting states with broken symmetry are observable,
either regular or chaotic. By focusing on the chaotic solutions, the so-called C2P state reported in Fig.~\ref{fig.1} (c),(f) and Fig.~\ref{fig.4} (a), we observe that in all the cases the oscillators of the two populations form a common cluster, characterized by
a common average frequency, plus a certain number of oscillators with larger frequencies 
(see Fig.~\ref{fig.4} (d)). These states resemble {\it imperfect chimeras} recently observed in experiments on coupled metronomes~\cite{kapitaniak2014} and in chains of rotators~\cite{jaros2015}.

The C2P states are characterized by a broken symmetry since the dynamics of the two populations takes place on different macroscopic chaotic attractors, as is clearly observable in Fig.~\ref{fig.4} (a) and (b). In particular, the differences in the oscillation amplitudes of $R^{(1)}$ and $R^{(2)}$  are due to the different number of oscillators contributing to the common cluster in the two populations~\cite{tanaka1997first,tanaka1997self, olmi2014}.
The multistability is also reflected in the associated Lyapunov spectra; three examples are shown in Fig.~\ref{fig.4} (c), their shapes are extremely different presenting different numbers of positive and negative LEs. However, a general result is that the oscillators contributing to the  chaotic dynamics are the ones out of the common cluster.
\red{ 
As shown in ~\cite{ginelli2011}, the contribution of each oscillator $i$ to the chaotic dynamics can be measured in terms of the corresponding squared component of the normalized maximal Lyapunov vector ${\cal T}^{(1)}$,  namely $\xi_i^{(\sigma)}(t) =[\delta\dot{\theta}_i^{(\sigma)}(t)]^2 + [\delta{\theta}_i^{(\sigma)}(t)]^2$ with $||{\cal T}^{(1)}(t) = 1||$. The time averaged components of the vector $\bar \xi_i$
are shown in Fig.~\ref{fig.4} (d), from where it is evident that the contribution $\bar \xi_i$ of the oscillators belonging to the common cluster is essentially negligible.}

\textit{Conclusions}.
We have shown that a simple model of coupled oscillators with inertia can reproduce the erratic behaviors observed in our experiment of two coupled populations of mechanical pendula. The presence of inertia is a distinctive ingredient to observe the emergence of chaotic regimes, like ICCs and C2Ps. The detailed characterization of the ICC dynamics reveals that its chaotic properties can be interpreted in the framework of fully coupled dissipative systems~\cite{takeuchi2011}. However, our study extends the validity of the results reported in~\cite{takeuchi2011} to networks with inhomogeneous coupling displaying intermittent dynamics. 
Together with the results reported in~\cite{wolfrum2011,wolfrum2011spectral} for 
a ring geometry, this clearly indicates that the stability properties of \emph{chaotic} chimeras strongly depend on the underlying network topology. 
It would be extremely challenging to investigate if the topology also influences 
the stability of non-chaotic chimeras.

Our dissipative model is quite remarkable, since it differs from a Hamiltonian model only by a constant dissipative term proportional to $1/m$ that vanishes in the limit of large inertia. This suggests that for sufficiently large $m$ dynamical properties characteristic of conservative models should be observable. Indeed, the Lyapunov spectrum exhibits a pairing rule~\cite{dressler1988} similar to that of Hamiltonian models. Furthermore, ICCs are metastable states, whose life-time diverges as $\simeq N^{1.6}$, in analogy to quasi-stationary states in the Hamiltonian mean-field model~\cite{campa2014_book}.

\begin{acknowledgments}
We thank D. Angulo-Garcia, S. Lepri, O.~E. Omel'chenko, A. Politi, S. Ruffo, K.~A. Takeuchi, M. Wolfrum for useful discussions and suggestions. We acknowledge partial financial support from 
the Italian Ministry of University and Research within the project CRISIS LAB PNR 2011-2013 (SO \& AT) and from the Human Frontier Science Program under a Cross Disciplinary Fellowship (ST). This work is part of the activity of the Marie Curie Initial  Training Network 'NETT' project \# 289146 financed by the European Commission (SO \& AT), and of the Dynamical Systems Interdisciplinary Network, University of Copenhagen (EAM).   
\end{acknowledgments}
 




\end{document}